\documentclass[preprint,12pt]{elsarticle}

\usepackage{lineno,hyperref}
\usepackage{fullpage}
\journal{ Elsevier}
\usepackage{amsmath,amssymb}
\usepackage{natbib}
\usepackage[below,section]{placeins} 
\usepackage{dsfont}
\usepackage{subcaption} 
\usepackage{rotating} 
\usepackage{mdframed}
\usepackage{booktabs, caption, makecell}

\usepackage{threeparttable}
\usepackage{siunitx }
\usepackage{hyperref}
\usepackage{pdflscape}
\hypersetup{
    colorlinks=true,
    linkcolor=blue,
    filecolor=magenta,      
    urlcolor=cyan,
}
 
\biboptions{authoryear}

\begin{document}

\begin{frontmatter}

\title{Nonparametric kernel estimation of Weibull-tail coefficient  in presence  of the right random censoring}


\author[rvt]{Justin USHIZE RUTIKANGA}
\ead{ushizerj@gmail.com}
\author[focal]{Aliou DIOP\footnote{corresponding author: Aliou DIOP}}
\ead{aliou.diop@ugb.edu.sn}
\address[rvt]{Universit\'e d'Abomey-Calavi, Institut de Math\'ematiques et de Sciences Physiques (\textbf{IMSP-UAC}), Porto-Novo, B\'enin}
\address[focal]{LERSTAD, Gaston Berger University, Saint-Louis, Senegal}

\begin{abstract}
In this paper, nonparametric estimation of the conditional Weibull-tail coefficient when the  variable of interest is  right random censored is addressed. A Weissman-type estimator of conditional extreme quantile is also proposed. 
In addition, a simulation study is conducted to assess the finite-sample behavior of the proposed estimators and a comparison with alternative strategies is provided. Finally, the practical  applicability of the  methodology is  presented  using  a real datasets of men suffering from a larynx cancer.
\end{abstract}

\end{frontmatter}


\section{ Introduction}\label{section1}
In statistics, the result of \cite{fisher_tippett_1928} on the laws of sample maximum, have been shown that the idea of extreme value theory has a big impact in analysis of the extreme events. Several application of rare events reappeared in different fields of real life situation  such as in non-life insurance, survival analysis and system or material reliability.\\

Due to  multiple sources of the information, some datasets of rare events are presented with  missing or incomplete information. Therefore, there have been numerous publications in the field of extreme value theory  studying the conditional extreme value index and the conditional extreme quantile under random right censoring, as presented in  \cite{stupfler2016estimating}. \\

According to the literature  several authors dealt with  problem of estimation of conditional extreme values  under censorship for heavy  tailed distributions; for instance see  \cite{ndao2014nonparametric, ndao2016nonparametric} among others. In their studies, they estimated the conditional extreme value index and  conditional extreme quantile  under random    right censorship for heavy tailed distributions in case  of fixed and/or random design. While \cite{stupfler2016estimating} generalized the work of \cite{ndao2015modelisation}, by  investigating whether its conditional distribution belongs to the Fr\'echet, Weibull or Gumbel  domain of attraction when the response
variable is right-censored.\\

However, most of studies were based on the assumption that the observed data come from heavy-tailed distributions (for both the censored and the censoring samples)  for instance \cite{diebolt2008bias, cabras_castellanos_2011, coles1996bayesian}. Nevertheless, in  \cite{stephenson2004bayesian}  the authors proposed  the Bayesian estimation extreme value index and extreme quantile  for the case  of uncensored  data by identifying three distinct types of extremal behaviour.
Besides, \cite{worms2014new,gomes2011estimation,matthys2004estimating} investigated  the estimation  of  conditional extreme value  index  and conditional  extreme quantile by considering non covariate information  as well as censored data are  taking into account. 
In some real life applications where the  rare events need to be studied, the tail heaviness of the conditional distribution
is not verified, particularly in the survival analysis, where the censored data are lifetimes of patients or of animals, or time-to-failure of systems or items. For example, in \cite{gomes2011estimation, worms2019estimation}, the authors have shown that the datasets of  men suffering from a larynx cancer did not exhibit a heavy right-tail.


This  study is motivated by the works of \cite{gardes2016estimation}, \cite{de2016kernel} and \cite{worms2019estimation}. In fact, \cite{gardes2016estimation}  estimated the conditional tail coefficient of Weibull-type distributions when functional covariate is available. \cite{de2016kernel} considered the estimation of the tail coefficient of a Weibull-type distribution in the presence of real random covariates. \cite{worms2019estimation} proposed an estimator of the Weibull-tail coefficient when the Weibull-tail
distribution of interest is censored from the right by another Weibull-tail distribution.

Our contribution in this paper is the  estimation of the conditional extreme quantile and conditional  tail coefficient for Weibull-type  distributions  under random right censoring, when real covariate is available.

This paper is organized as follows. Section \ref{section2} is devoted to the theoretical framework, while Section \ref{section3} dealt with  the construction of our proposed  estimators. 
The finite sample behaviour of the proposed estimators is  examined in Section \ref{section4} in a simulation study. 
A real data application illustrate the use of our estimators in Section \ref{section5}.
Finally, Section \ref{section6} concludes the paper and gives some perspectives.
\section{Framework}\label{section2} 

Let $ (X_i,Y_i) \quad i=1,\cdots, n$ be the independent  copies of the   random pairs $(X,Y)$, where $Y$ is  positive real random variable and $X$ be a real random variable, $X \in \mathbb{R}^d,d\geq 1 $. We assume that $Y$ can be right-censored by a non-negative random variable $C$.

Let us consider the observation  of sample of $n$ independent  triplets $(X_i,Z_i,\delta_i)_{1\leqslant i \leqslant n}$ where  $ Z_i= \min(Y_i,C_i)$ and $\delta_i=\mathds{1}_{\{Y_i\leqslant C_i\}}$ for $i=1\cdots,n$ where $\mathds{1}_{\{A\}}$ is the indicator function of the event A.\\

 By considering that the  i.i.d sample $(Y_i)_{i \leqslant n}$ and $(C_i)_{i \leqslant n}$   respectively  has continuous distribution  function. Let $F$ and $G$ be  distribution function  of  the variable of interest $Y$ and the censoring variable $C$ respectively. The variable $Y$ and $C$  are supposed  to be independent as adopted  in  \citep{ndao2014nonparametric}.\\
 
In this paper, we will use $ Z_{1,n}\leqslant \cdots \leqslant Z_{n,n}$ as the ordered statistics associated to the observed  sample  and $(\delta_1, \cdots , \delta_n)$ the corresponding observed non-censoring indicators.\\

 
The main goal of this work  is to investigate  the behavior  of the right tail of the conditional distribution of $F$ given $X=x$.
Suppose that 
\begin{eqnarray}
\bar{F}(y|x)=1-F(y|x)=\exp(-\Lambda_F(y|x))\\
\bar{G}(c|x)=1-G(c|x)=\exp(-\Lambda_G(c|x)),
\end{eqnarray}
where $ \Lambda_F(\cdot|x)$ and $ \Lambda_G(\cdot|x)$ are   conditional cumulative  hazard  function  of $Y$ and $C$ given $X=x$ respectively. In this work  we assume that  both  the censored and censoring  variables  have the  Weibull-tail type  distribution.\\
Let us suppose  that the conditional cumulative distribution  function $F(\cdot|x)$ has  a Weibull tail if the following condition in \eqref{Eqfbar}  holds and there exists  a function $\gamma_Y(x)>0$ such that  for all $\lambda>0$

\begin{eqnarray}\label{Eqfbar}
\lim_{y\rightarrow \infty}\frac{\log(1-F(\lambda y|x))}{\log(1-F(y|x))}= \lambda^{1/\gamma_Y(x)}
\end{eqnarray}
or can be written  as
\begin{eqnarray*}
\lim_{y\rightarrow \infty}\frac{\Lambda_F(\lambda y|x)}{\Lambda_F(y|x)}= \lambda^{1/\gamma_Y(x)}
\end{eqnarray*}
where $\Lambda_F(\cdot|x)$ is the cumulative hazard function of random variable $Y$ given $X=x$.
 The parameter $\gamma_Y(x)$  is referred  as the  conditional Weibull tail coefficient.

There exists  some positive  parameters $\gamma_Y(x)$ and $\gamma_C(x)$ and some slowly  varying function at infinity $\ell_F(\cdot|x)$ and $\ell_G(\cdot|x)$ such that  for every $y$ and $c$
\begin{eqnarray}
\Lambda_F(y|x)=y^{1/\gamma_Y(x)}\ell_F(y|x) \quad \textit{and}\quad \Lambda_G(c|x)=y^{1/\gamma_c(x)}\ell_G(c|x).
\end{eqnarray}
Then let  $H$ be the cumulative  distribution  function  of the observed  variable $Z$ and
\begin{eqnarray}
\bar{H}(y|x)=1-H(y|x)=P(Z>y|x),
\end{eqnarray} 
 the independence of the samples $Y$ and $C$ given $X=x$ allows us to write  
 \begin{eqnarray}
 \bar{H}(y|x)=\bar{F}(y|x)\bar{G}(y|x)=\exp(-\Lambda_H(y|x)), 
\end{eqnarray} 
 with 
\begin{eqnarray}
\Lambda_H(y|x)= \Lambda_F(y|x)+\Lambda_G(y|x)=y^{1/\gamma_Y(x)}\ell_F(y|x)+y^{1/\gamma_c(x)}\ell_G(y|x)=y^{1/\gamma_Z(x)}\ell_H(y|x)
\end{eqnarray}
$\Lambda_H(\cdot|X=x)$ of $Z$ given $X=x$ is also a regularly varying  function at infinity with index $1/\gamma_Z(x)$ 
where $\gamma_Z(x)=\min(\gamma_Y(x),\gamma_C(x))$.\\

The associated  conditional quantile is defined as follows
\begin{eqnarray}
q(\alpha|x)=\bar{H}^{-1}(\alpha|x)=\Lambda_H^{-1}(\log(1/\alpha)|x)
\end{eqnarray}
for all $\alpha \in (0,1)$. Here, we are dealing with the Weibull  tail distribution, then $\Lambda_H(\cdot|x)$ is a regularly  varying function at infinity with index $1/\gamma_Z(x)$:
\begin{eqnarray}
\lim_{y \rightarrow \infty} \frac{\Lambda_H(ty|x)}{\Lambda_H(y|x)}=t^{1/\gamma_Z(x)} \quad \forall t \geq 0
\end{eqnarray}
with $\gamma_Z(x)$ is an unknown positive function of the covariate $x \in \mathbb{R}^d$.

This implies that $ \Lambda_H^{-1}(\cdot|x)$ is also a regularly varying function at infinity  with index $\gamma_Z(x)$ and  the below relation  hold:
\begin{eqnarray}
q(e^{-y}|x)=\Lambda_H^{-1}(y|x)=y^{\gamma_Z(x)}\ell(y|x)
\end{eqnarray}
where $\ell(\cdot|x)$ is a regularly varying function at infinity  such that 
\begin{eqnarray}
\lim_{y \rightarrow \infty} \frac{\ell(ty|x)}{\ell(y)}=1 \quad \forall t\geq 0.
\end{eqnarray}

\section{Construction of the estimators}\label{section3}
In literature there exists  several estimators of the  tail coefficient of type of the Weibull-tail  distribution. The first estimator was proposed by \cite{berred1991record} where  the estimator was  constructed based on the  record values.  The most used one  was proposed by \cite{beirlant1996practical}, it is constructed based on the definition of the  quantile function of distribution of the  tail of type  Weibull.
Later \cite{beirlant2004estimation} introduce  other estimator  which was based on the  logarithmic of the threshold excess of the $k_n$ highest ordered statistics  in the sample. 
\begin{eqnarray}
\log Y_{n-i+1,n} -\log Y_{n-k_n+1,n}
\end{eqnarray}
The most popular is the estimator  proposed by \cite{goegebeur2014nonparametric,gardes2012functional}, this  estimator  was derived  based on the log spacings 
\begin{eqnarray}
\log Y_{n-i+1,n} -\log Y_{n-i,n}
\end{eqnarray}

Let $\alpha$ and $\beta$ be a real numbers closed to zero and define
\begin{eqnarray}\label{q1}
q(\alpha)=\log(1/\alpha)^{\gamma}\ell(\log(1/\alpha))
\end{eqnarray}
\begin{eqnarray}\label{q2}
q(\beta)=\log(1/\beta)^{\gamma}\ell(\log(1/\beta)).
\end{eqnarray}
By taking logarithmic for Equation \eqref{q1} and \eqref{q2} then subtract Equation \eqref{q1} into \eqref{q2} , we obtained
\begin{eqnarray*}
\log(q(\beta))-\log(q(\alpha))&=&\gamma \log\log(1/\beta)+\log(\ell(\log(1/\beta)))-\gamma \log\log(1/\alpha)-\log(\ell(\log(1/\alpha)))\\
&=&\gamma\left[\log\log(1/\beta)-\log\log(1/\alpha)\right] +\log\left[\frac{\ell(\log(1/\beta))}{\ell(\log(1/\alpha))}\right]\\
\end{eqnarray*}
Since $\ell$ is a slowly varying function at infinity and  consider $\alpha=i/n$ and $\beta=k_n/n$, $\log_2(\cdot)=\log\log(\cdot)$  therefore,
\begin{eqnarray}
\log(q(\beta))-\log(q(\alpha))& \approx & \gamma\left[ \log_2(n/i)-\log_2(n/k_n)\right]\\
\gamma &\approx & \frac{\log(q(\beta))-\log(q(\alpha))}{ \log_2(n/i)-\log_2(n/k_n)}
\end{eqnarray}
then the estimator of $\gamma$ is given by
\begin{eqnarray}
\hat{\gamma}=\sum_{i=1}^{k_n}\left[\log(Y_{n-i+1,n})-\log(Y_{n-k_n+1,n})\right]/\sum_{i=1}^{k_n}\left[ \log_2(n/i)-\log_2(n/k_n)\right]
\end{eqnarray}

By controlling  the presence  of the covariate,  we adopted the estimator  proposed by \cite{goegebeur2014local} for a Hill's type estimator for conditional extreme value index expressed as follows. Let $y_n$ be a non-random  sequence such that $ y_n \rightarrow \infty$ as $n \rightarrow \infty$\\
\begin{eqnarray}\label{EqHest1}
\hat{\gamma}_Y^{(complete)}(x)= \frac{\sum_{i=1}^n K((x-X_i)/h)\left[ \log(Y_i)-\log(y_n) \right]\mathds{1}_{\{Y_i\geqslant y_n\}}}{\sum_{i=1}^n K((x-X_i)/h)\left[ \log_2(n/i)-\log_2(n/k_n)\right]\mathds{1}_{\{Y_i\geqslant y_n\}}}.
\end{eqnarray}
where $K$ is a kernel density  function and  $h$ a positive sequence of non-random  bandwidth  such that  $h$ goes  to zero as $n \rightarrow \infty$.\\

The estimator \eqref{EqHest1} is not consistent for $\gamma_Y(x)$ if it is directly applied to the censored sample $(X_i,\delta_i,Z_i), i=1,\cdots, n$.
Indeed, under appropriate regularity assumptions, in our censored framework, we propose the following estimator
\begin{eqnarray}\label{EqHest2}
\hat{\gamma}_Y(x)=   \frac{\sum_{i=1}^n K((x-X_i)/h)\left[ \log(Z_i)-\log(y_n) \right]\mathds{1}_{\{Z_i\geqslant y_n\}}}{\sum_{i=1}^n K((x-X_i)/h)\left[ \log\hat{\Lambda}_{n,F}(Z_i|x)-\log\hat{\Lambda}_{n,F}(y_n|x)\right]\mathds{1}_{\{Z_i\geqslant y_n\}}}
\end{eqnarray}

 where
  \begin{eqnarray}\label{EqHazard}
\hat{\Lambda}_{n,F}(y_n|x)= \sum_{i:Z_{(i)}\leq y_n}^n \frac{B_{i}(x)\mathds{1}_{\{Z_i > y_n,\delta_i=1\}}}{1-\sum_{j=1}^{i-1} B_{j}(x)\mathds{1}_{\{Zj\leqslant y_n\}}}.
\end{eqnarray}
$\hat{\Lambda}_{n,F}(\cdot|x)$ is a nonparametric estimator of $\Lambda_{F}(\cdot|x)$ which  is known as the Beran estimator of conditional cumulative hazard function. However, for sake of simplicity in our simulation we will use $\hat{\Lambda}_{n,F}(\cdot|x)=-\log \widehat{\bar{F}}_n(\cdot|x)$, with  $\widehat{\bar{F}}_n$ is the kernel conditional Kaplan-Meier estimator adapted in \cite{ndao2016nonparametric} which depends on parameter $h$. 
We can also use
 \begin{eqnarray}\label{EqHazard}
\hat{\Lambda}_{n,F}(y_n|x)&=& \int_0^{y_n}\frac{dH_{1n}(s|x)}{1-H_n(s|x)} \qquad\textit{with}\\ \nonumber
H_n(y_n|x)&=& \sum_{i=1}^nB_{i}(x)\mathds{1}_{\{Zj\leqslant y_n\}}\\ \nonumber
H_{1n}(y_n|x)&=& \sum_{i=1}^nB_{i}(x)\mathds{1}_{\{Zj\leqslant y_n,\delta=1\}}\\ \nonumber
\end{eqnarray}
and 

\begin{eqnarray}
B_i(x) =\frac{K(h^{-1}(x-X_i))}{\sum^n_{j=1}K(h^{-1}(x-X_j))}.
\end{eqnarray} 
 with $B_{i}(x)$ is well known the Nadaraya Watson weighted \citep{nadaraya1964estimating,watson1964smooth}.\\

Using the abovementioned estimators $\hat{\gamma}_{Y}(x)$ of $\gamma_Y(x)$, let us now consider the  estimation of extreme quantiles for Weibull-tail  under right random censored  data. For any given small probability $\alpha_n$, we can now  adapt the classical estimator of $q(\alpha_n|x)=F^{\leftarrow}(\alpha_n|x)$  proposed in \cite{worms2019estimation} as follows:

\begin{eqnarray}\label{EqWquantile}
\hat{q}_w(\alpha_n|x) = y_n\left[ \frac{-\log(\alpha_n)}{\hat{\Lambda}_{n,F}(y_n|x)}\right]^{\hat{\gamma}_{Y}(x) }.
\end{eqnarray} 

\section{ Simulation studies}\label{section4}

\subsection{Simulation design}\label{design}
In this section, the main purpose is to illustrate  our methodology with a simulation experiment. The finite sample performances of our proposed estimators  $\hat{\gamma}_Y(x)$ and $\hat{q}_w(\alpha_n|x)$ (for small $\alpha_n$) in terms of mean absolute errors (MAE)  and mean squared error (MSE) are performed.\\

\begin{figure}[ht!]
\centering
  \includegraphics[width=14cm,height=10cm]{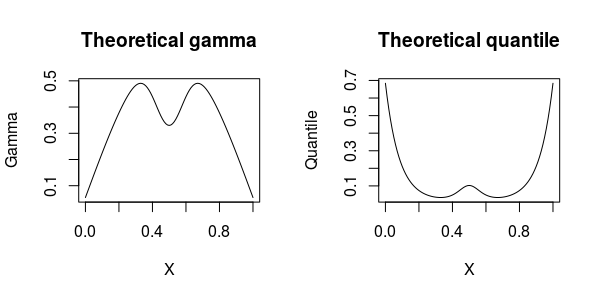}
\caption{Pattern of $\gamma_Y(\cdot)$ and $q(1/1000|\cdot)$ on $[0,1]$ .} 
\label{fig:1}
\end{figure}

\noindent To  achieve our goal, we consider the simulation of  $N=500$ replications  of the sample of size $n(n = 500, 300, 100)$  of  random triplets $(Z_i,\;\delta_i, \; X_i)$, where $Z_i=\min(Y_i,C_i)$ and $X_i$ is uniformly distributed on $[0,1]$.
 The conditional  distribution of $Y$ given $X=x$  is Weibull distribution  with  conditional  cumulative  distribution function $ 1-\exp(-y^{1/\gamma_Y(x)})$. The parameter $\gamma_Y(x)$ is given in the following equation 
 \begin{eqnarray}\label{gamma}
\gamma_Y(x)=0.5(0.1+\sin(\pi x))(1.1-0.5\exp(-64(x-0.5)^2)). 
\end{eqnarray}
 Figure \ref{fig:1} illustrates  the pattern of the theoretical  value of $\gamma_Y(\cdot)$ and $q(1/1000|\cdot)$ on $[0,1]$.

\FloatBarrier
 The conditional  distribution  of $C$ given $X=x$  is also  Weibull and its parameter $\gamma_C(x)$ is chosen to yield three scenarios such  as $\gamma_Y(x)<\gamma_C(x),\gamma_Y(x)=\gamma_C(x) $ and $\gamma_Y(x)>\gamma_C(x)$, corresponding to different intensities of censoring in the tail. For each of the $N$ samples, we estimate $\gamma_Y(\cdot)$ at different value of $x=(0.1,0.2,0.3,0.4,0.5,0.6,0.7,0.8,0.9)$ for example  $x=0.5 \quad (\gamma_Y(0.5)=0.33 )$ using  our estimator  presented in Equation \eqref{EqHest2}.\\
 
\noindent In  this simulation experiment, we are interest to  estimate $\gamma_Y(x)$ using the estimators $\hat{\gamma}_Y^{(complete)}(x)$, $\hat{\gamma}_Z^{(complete)}(x)$ and $\hat{\gamma}_Y(x)$, with an asymmetric linear kernel  defined as $K(u)= (1.9-1.8 u)\mathds{1}_{u\in [-1,1]}$.\\
Our proposed estimators depend on the  bandwidth parameter $h_n=h$ which is chosen using a data-driven method since it does not require any prior knowledge about the function $\gamma_Y(x)$. However, the  bandwidth parameter $h$ is chosen using the cross-validation  method  which were implemented in \cite{gardes2012functional}.
\begin{eqnarray}
h^{opt}=\arg\min _{h_n \in \mathcal{H}}\sum_{i=1}^n\sum_{j=1}^n \left( \mathbb{\textbf{1}}_{\{Z_i>Z_j\}}- \widehat{\bar{F}}_{n,-i}(Z_j|x_i) \right)^2,\nonumber
\end{eqnarray}
where $\mathcal{H}$ is a grid of values for $h_n$
 and  $\widehat{\bar{F}}_{n,-i}$ is the kernel conditional Kaplan-Meier estimator  adapted in \cite{ndao2016nonparametric}, which depends on parameter $h_n$.\\

 \noindent In case the bandwidth has already been selected, we adopt  the  method used in \cite{ndao2016nonparametric}  to choose the  threshold  excess $k_x =k$ and  described as follows: 
 \begin{enumerate}
\item we compute the estimate  $\hat{\gamma}_Y(x)$ with $ k=1,\cdots,n-1$,
\item we form several successive "blocks" of estimates $\hat{\gamma}_Y(x)$ (one block for $k\in \{1,\ldots,10\}$, a second block for $k\in \{11,\ldots,20\}$ and so on),
\item we calculate the standard deviation of the estimates within each block,
\item we determine the $k$-value to be used (thereafter denoted by  $k^*$) from the block with minimal standard deviation. Precisely,
we take the middle value of the $k$-values in the block (see \cite{ndao2016nonparametric, goegebeur2014local})
\end{enumerate}

\noindent In simulation experiences, the choice of kernel  density function does not shown any impact in performance of our proposed estimators. By rerun our experiments  with other kernel density function 
 for example of a bi-quadratic kernel defined as $K(u)= \frac{15}{16}(1- u^2)^2 \mathds{1}_{u \in[-1,1]}$ and there is no impact in the results.\\
 
\noindent  Regarding the estimation of the upper extreme quantile, we  performed  different experiment by examining  the behaviors of $\hat{q}_w(1-\alpha_n|x) $ with respect to  the value of $\hat{\gamma}_Y^{(complete)}(x)$, $\hat{\gamma}_Z^{(complete)}(x)$ and $\hat{\gamma}_Y(x)$ where $\alpha_n=1/1000$.
\subsection{Results}

\noindent Table \ref{tab:1} and Table \ref{tab:2} give an overview of the performances of our estimators  of the conditional  Weibull-tail coefficient $\gamma_Y(x)$ and conditional extreme quantile $q(\alpha_n|x)$ for small  probability $\alpha_n$.  Based on the value of empirical Mean Squared Error (MSE) and  empirical Mean Absolute Error (MAE) over the $N$ estimates, we access the accuracy  of our proposed estimators based on the different censoring  intensities. 

To  demonstrate  the effectiveness of our estimator $\hat{\gamma}_Y(x)$ presented in Equation \eqref{EqHest2}, we also compare it with $\hat{\gamma}^{(comp)}_Y(x)$ defined in Equation \eqref{EqHest1}, by considering that the sample $Y$ is observed  but practically is wrong, since  we can not observed  a sample $Y$ in censored framework. Again  we make  the comparison  with $\hat{\gamma}^{(comp)}_Z(x)$ which is the same expression as $\hat{\gamma}^{(comp)}_Y(x)$ but applied to the observed sample $Z$ instead  to sample $Y$. For each experiment, we considered three scenario 
$\gamma_Y(x)<\gamma_C(x),\gamma_Y(x)=\gamma_C(x)$ and $\gamma_Y(x)>\gamma_C(x)$ .

Table \ref{tab:1} illustrate  the different value of empirical MSE and empirical MAE of our estimators of Weibull tail coefficient $\gamma_Y(x)$ at different sample size with respect to the different censoring intensities respectively. As expected the  simulation  study show that 
an estimator $\hat{\gamma}^{(comp)}_Z(x)$ not adapted to censoring framework yields inaccurate results, even in the case $\gamma_Y(x)<\gamma_C(x)$ , where  $\hat{\gamma}^{(comp)}_Z(x)$ is consistent for estimating $\gamma_Y(x)$.

As illustrated from Table \ref{tab:1}, the Hill's kernel version estimator under censorship proposed
estimator presented  in Equation \eqref{EqHest2} of $\gamma_Z(x)$ shows to be well performed in almost simulation cases. As result, it performs quite better on the scenario $\gamma_Y(x)<\gamma_C(x)$ for large enough sample size and its quality becomes worst as $\gamma_Y(x)>\gamma_C(x)$ and sample
size decreases. 
 
\begin{figure}[ht!]
\centering
\includegraphics[width=15cm,height=20cm]{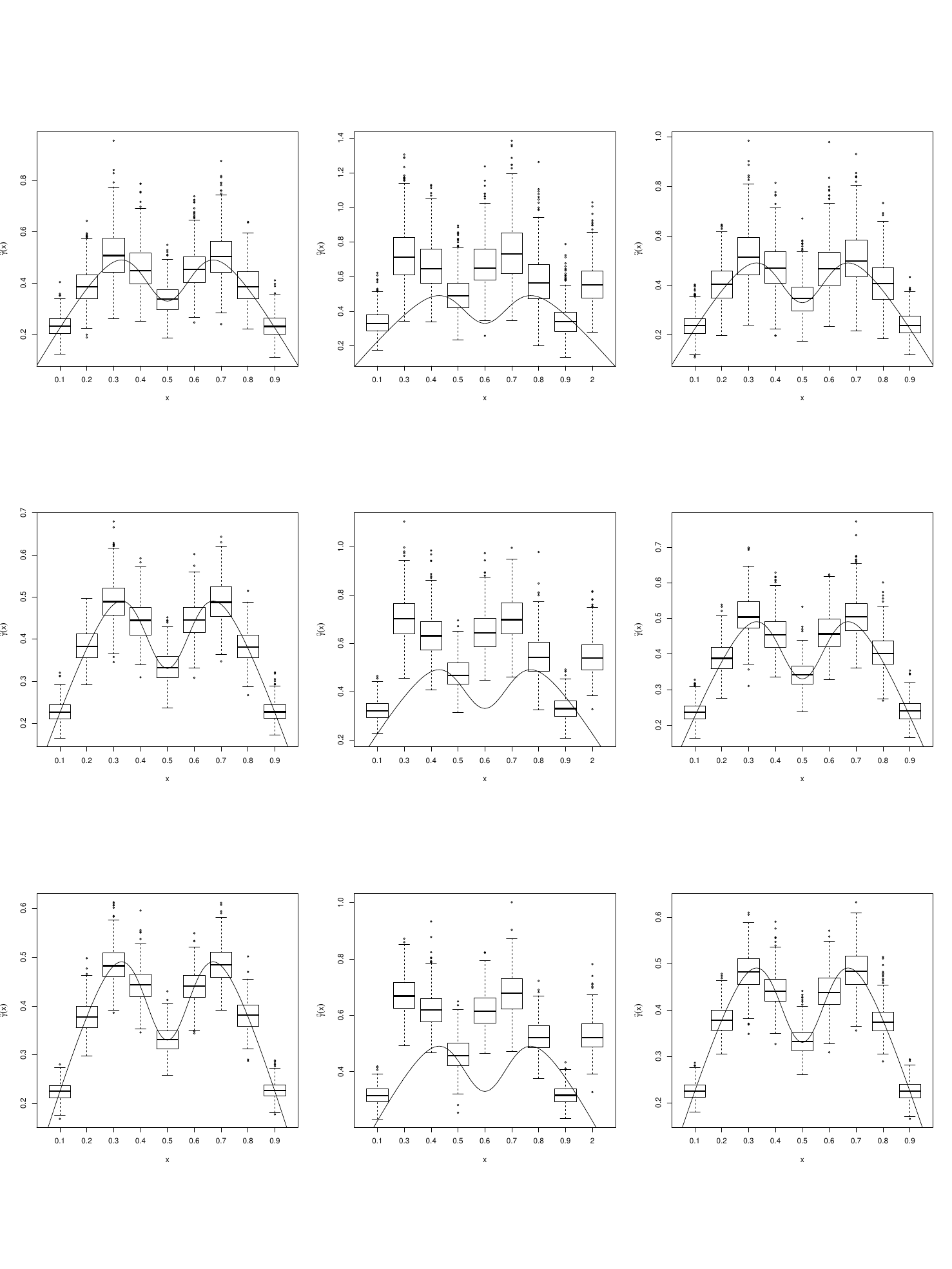}
\caption{Pattern simulation for  $N=500$ of estimates of $\gamma_Y(\cdot)$ on $[0,1]$ with $n=100$ first line, $n=300$ second line, $n=500$. Left $\hat{\gamma}_Y(x)$, center $\hat{\gamma}^{(comp)}_Z(x)$ and right $\hat{\gamma}^{(comp)}_Y(x)$ where $\gamma_Y< \gamma_C$.} 
\label{fig:2}
\end{figure}

\begin{figure}[ht!]
\centering
   \includegraphics[width=15cm,height=20cm]{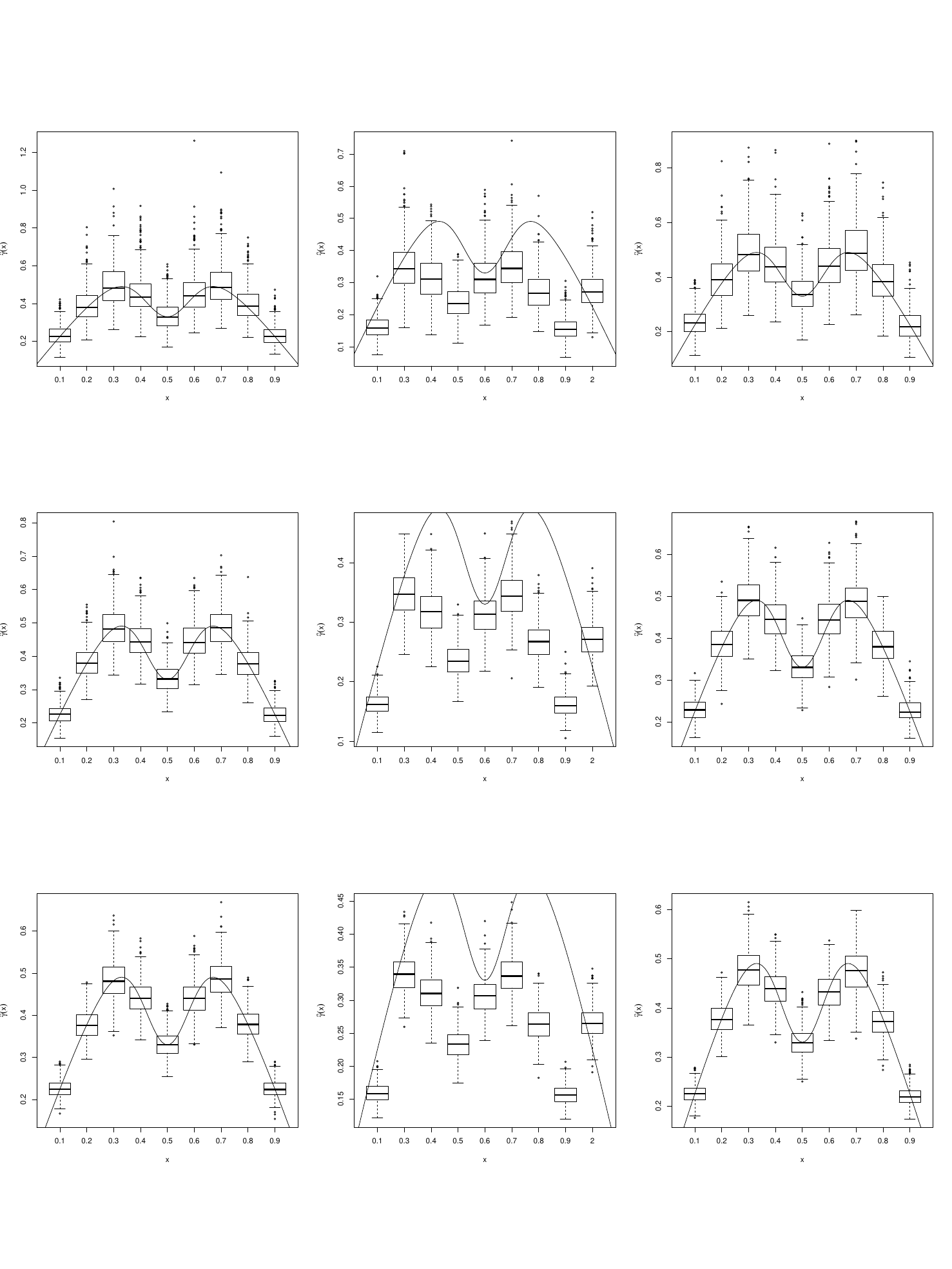}
\caption{Pattern simulation for  $N=500$ of estimates of $\gamma_Y(\cdot)$ on $[0,1]$ with $n=100$ first line, $n=300$ second line, $n=500$. Left $\hat{\gamma}_Y(x)$, center $\hat{\gamma}^{(comp)}_Z(x)$ and right $\hat{\gamma}^{(comp)}_Y(x)$ where $\gamma_Y> \gamma_C$.} 
\label{fig:3}
\end{figure}

\begin{figure}[ht!]
\centering
  \includegraphics[width=15cm,height=20cm]{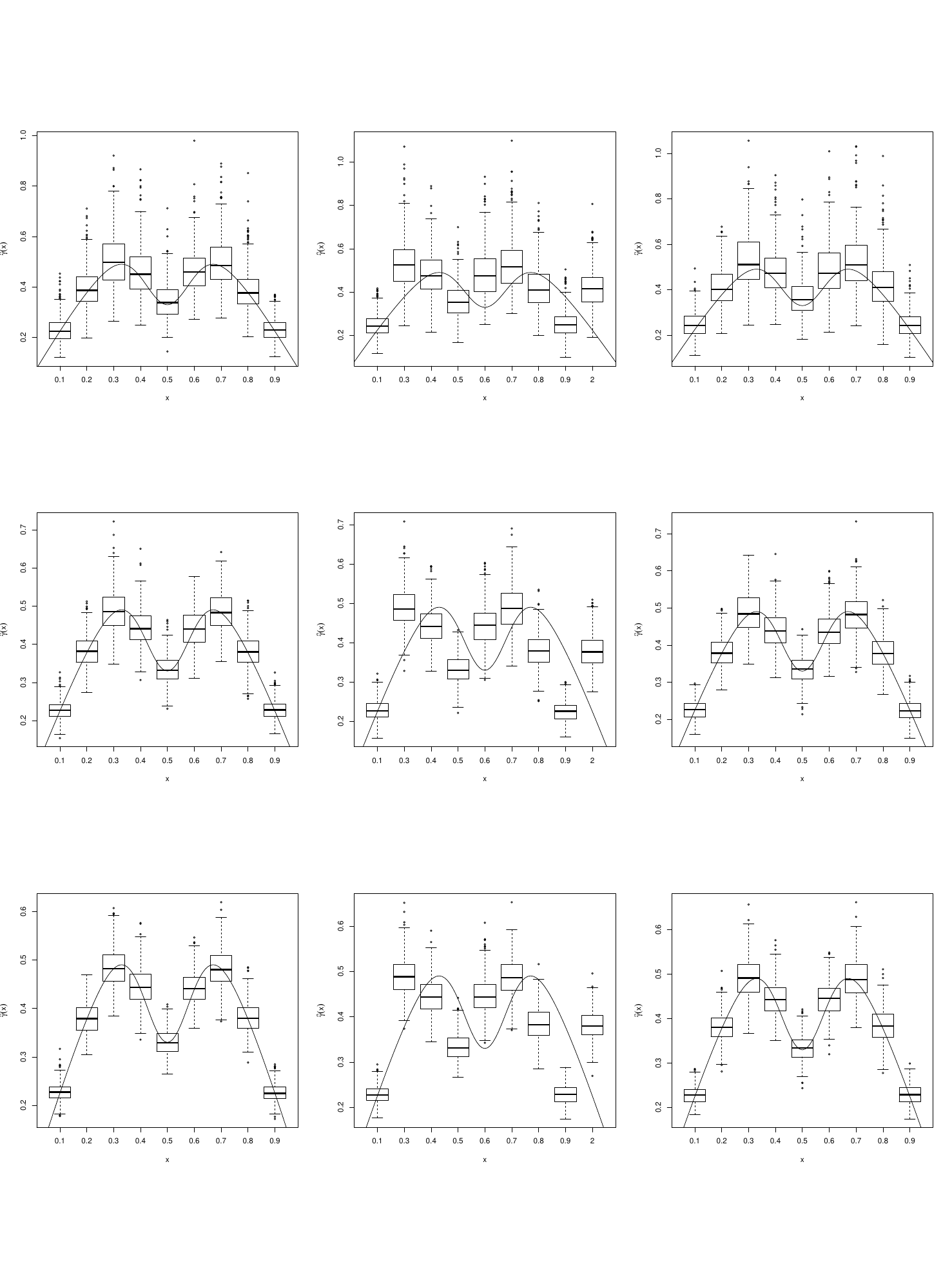}
\caption{Pattern simulation for  $N=500$ of estimates of $\gamma_Y(\cdot)$ on $[0,1]$ with $n=100$ first line, $n=300$ second line, $n=500$. Left $\hat{\gamma}_Y(x)$, center $\hat{\gamma}^{(comp)}_Z(x)$ and right $\hat{\gamma}^{(comp)}_Y(x)$ where $\gamma_Y= \gamma_C$.} 
\label{fig:4}
\end{figure}

According to  the Boxplot  of the estimators  of $ \gamma_Y(x)$  for $N=500$ random sample  of sample size $n=100, 300, 500$  presented in Figure \ref{fig:2}, \ref{fig:3}  and \ref{fig:4} for the case $\gamma_Y(x)<\gamma_C(x)$,  $\gamma_Y(x)=\gamma_C(x)$ and $\gamma_Y(x)>\gamma_C(x)$  respectively. As expected, the figures show that  the proposed  estimator for $ \gamma_Y(x)$ in framework of censorship is well  performed  for all scenarios  as  the sample  size  is large enough.

In Table \ref{tab:2}, we present  different value of empirical MSE and empirical MAE of our
estimators of extreme conditional quantile  $q(\cdot|x)$ at different sample size increases with respect to the different censoring intensities respectively. As expected the  simulation  study show that 
an estimator correspondent  to (here $\hat{\gamma}^{(comp)}_Z(x)$) not adapted to censoring yields inaccurate results, even in the case $\gamma_Y(x)<\gamma_C(x)$.

As illustrated from Table \ref{tab:2}, Weissman quantile  estimator under censorship proposed
estimator presented  in Equation \eqref{EqWquantile} of $q(\cdot|x)$ shows to be well performed in almost simulation cases. As result, it performs quite better on the scenario $\gamma_Y(x)<\gamma_C(x)(x)$ for large enough sample size and its quality becomes worst as $\gamma_Y(x)>\gamma_C(x)$ and sample
size decreases. 
 
\begin{figure}[ht!]
\centering
  \includegraphics[width=15cm,height=20cm]{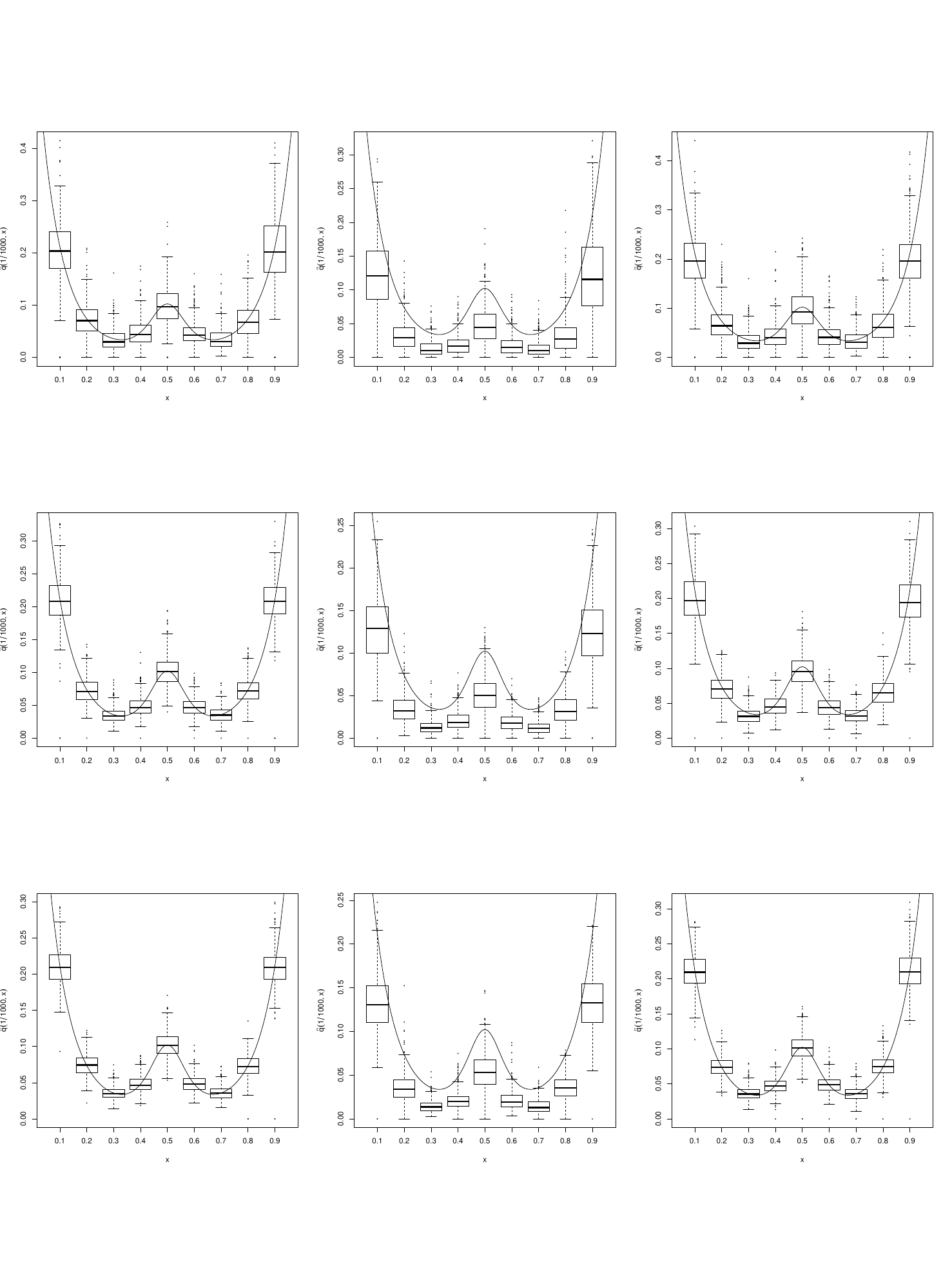}
\caption{Pattern simulation for  $N=500$ of estimates of $q(1/1000|\cdot)$ on $[0,1]$ with $n=100$ first line, $n=300$ second line, $n=500$. For each $\hat{q}_w(1/1000|\cdot)$  corresponding to left by $\hat{\gamma}_Y(x)$, center by $\hat{\gamma}^{(comp)}_Z(x)$ and right by $\hat{\gamma}^{(comp)}_Y(x)$ where $\gamma_Y< \gamma_C$.} 
\label{fig:5}
\end{figure}

\begin{figure}[ht!]
\centering
  \includegraphics[width=15cm,height=20cm]{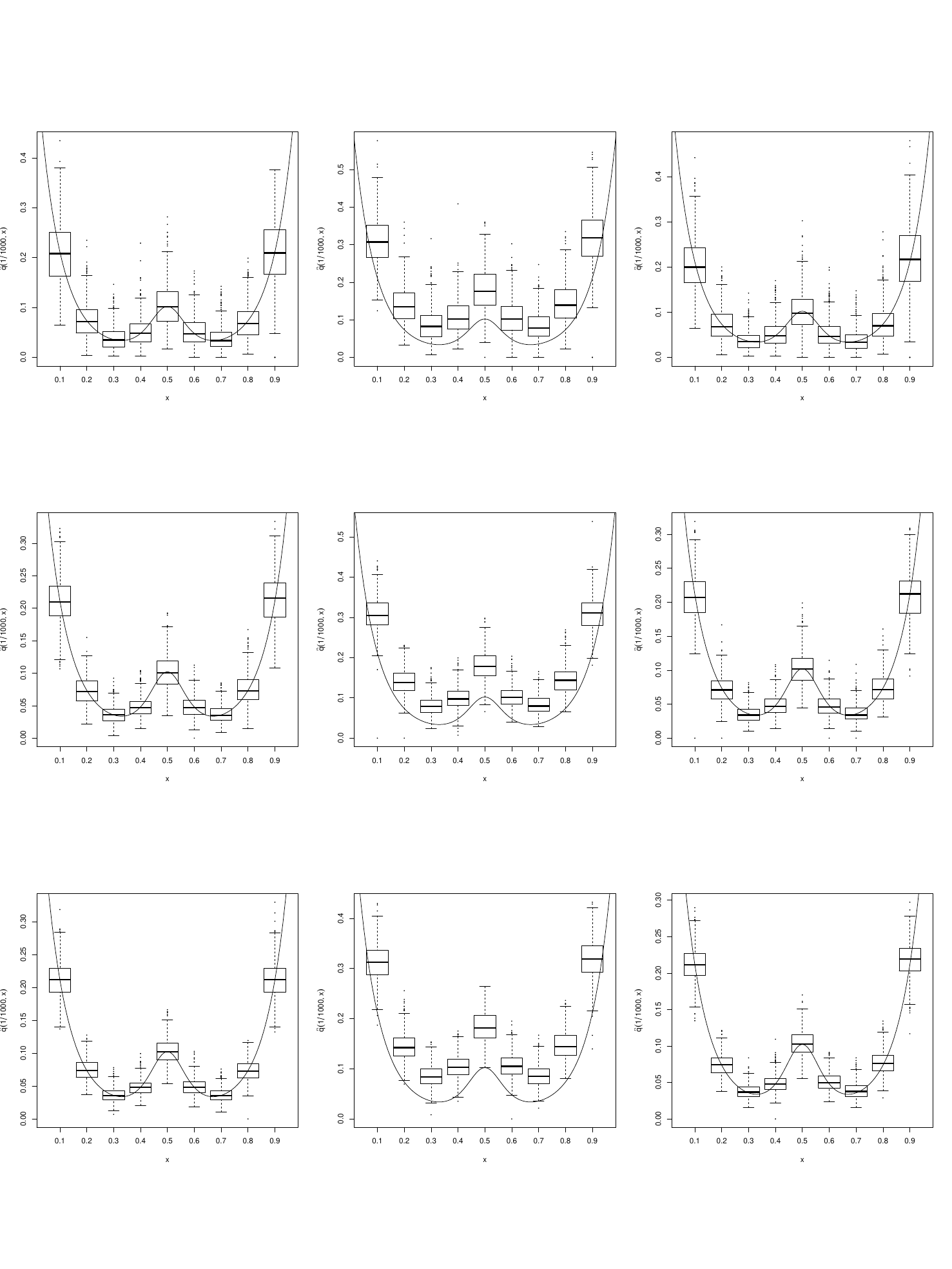}
\caption{Pattern simulation for  $N=500$ of estimates of $q(1/1000|\cdot)$ on $[0,1]$ with $n=100$ first line, $n=300$ second line, $n=500$.  For each $\hat{q}_w(1/1000|\cdot)$  corresponding to left  by $\hat{\gamma}_Y(x)$, center  by $\hat{\gamma}^{(comp)}_Z(x)$ and right by  $\hat{\gamma}^{(comp)}_Y(x)$ where $\gamma_Y> \gamma_C$.} 
\label{fig:6}
\end{figure}

\begin{figure}[ht!]
\centering
  \includegraphics[width=15cm,height=20cm]{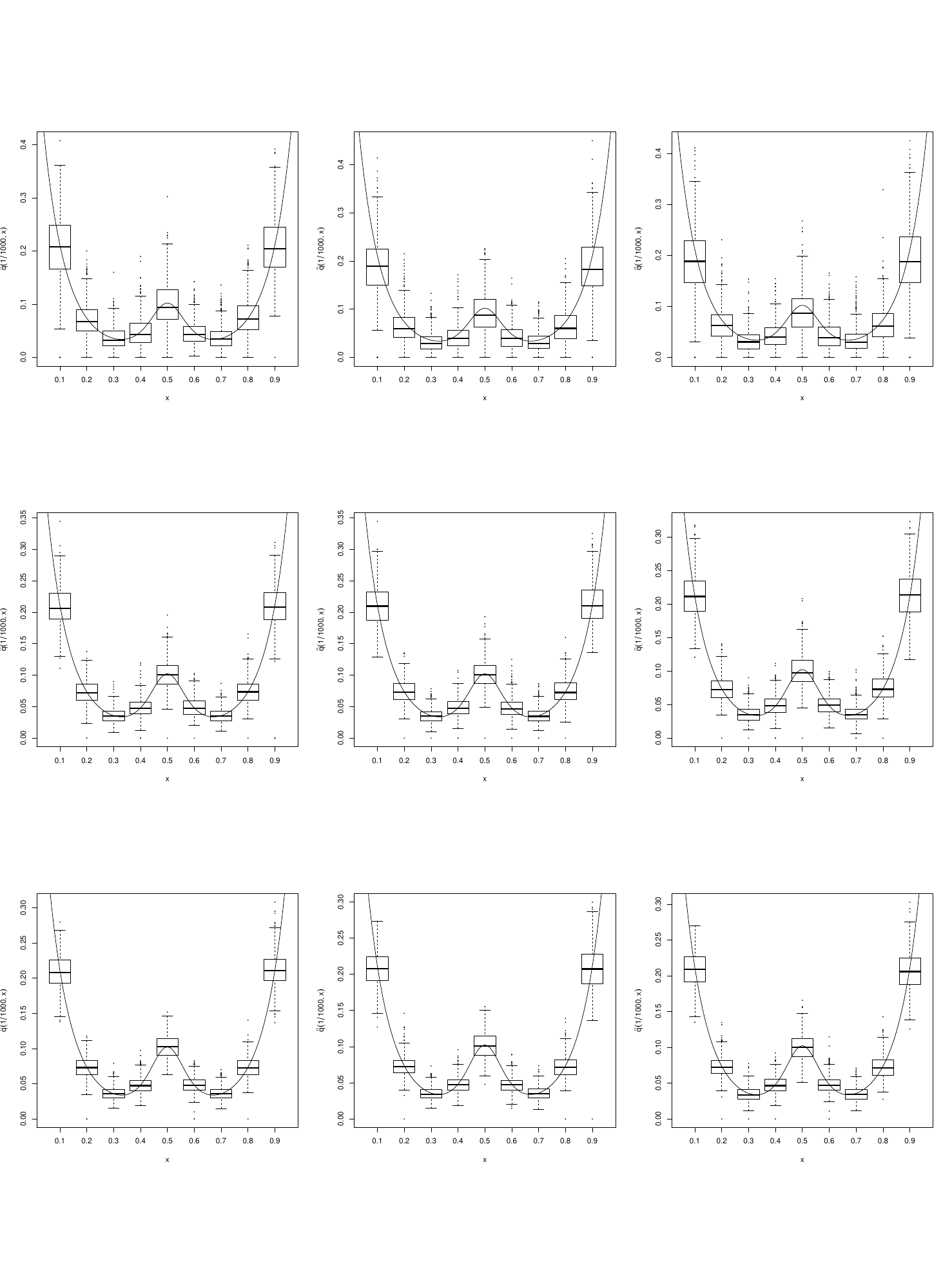}
\caption{Pattern simulation for  $N=500$ of estimates of $q(1/1000|\cdot)$ on $[0,1]$ with $n=100$ first line, $n=300$ second line, $n=500$. For each $\hat{q}_w(1/1000|\cdot)$  corresponding to left by $\hat{\gamma}_Y(x)$, center by $\hat{\gamma}^{(comp)}_Z(x)$ and right by $\hat{\gamma}^{(comp)}_Y(x)$ where $\gamma_Y= \gamma_C$.} 
\label{fig:7}
\end{figure}

According to  the Boxplot  of the estimators  of $ q(\cdot|x)$  for $N=500$ random sample  of sample size $n=100,300, 500$  presented in Figures \ref{fig:5}, \ref{fig:6}  and \ref{fig:7} for the case $\gamma_Y(x)<\gamma_C(x)$,  $\gamma_Y(x)=\gamma_C(x)$ and $\gamma_Y(x)>\gamma_C(x)$  respectively. The result shows that the proposed  estimator for $ q(\cdot|x)$ in framework of censorship is well  performed  for all scenarios  as  the sample  size  is large enough.

\begin{sidewaystable}[ht!]
\caption{simulation results for the estimator of $\gamma_Y(x)$: empirical MSE, empirical MAE for $N = 500$ replications.}
  \label{tab:1}
\centering
\scalebox{.65}{\begin{tabular}{c|c|cc|cc|cc|cc|cc|cc|cc|cc|cc}
\hline
&estimator&\multicolumn{2}{|c|}{ $\gamma_Y(0.1)=0.2249$ }&\multicolumn{2}{|c|}{  $\gamma_Y(0.2)=0.3777$  }&\multicolumn{2}{|c}{ $\gamma_Y(0.3)=0.4823$}&\multicolumn{2}{|c|}{ $\gamma_Y(0.4)=0.4395$ }&\multicolumn{2}{|c|}{  $\gamma_Y(0.5)=0.3300$  }&\multicolumn{2}{|c}{ $\gamma_Y(0.6)=0.4395$}&\multicolumn{2}{|c|}{ $\gamma_Y(0.7)=0.4823$ }&\multicolumn{2}{|c|}{  $\gamma_Y(0.8)=0.3777$  }&\multicolumn{2}{|c}{ $\gamma_Y(0.9)=0.2249$}\\
\hline
$n$&&MSE&MAE&MSE&MAE&MSE&MAE&MSE&MAE&MSE&MAE&MSE&MAE&MSE&MAE&MSE&MAE&MSE&MAE\\
\hline
\multicolumn{20}{c}{ For $\gamma_Y<\gamma_C$ }\\
\hline
&$\hat{\gamma}^{(comp)}_Y(x)$&0.0019& 0.0340&0.0053& 0.0565& 0.0086& 0.0713&0.0081& 0.0701& 0.0047& 0.0529&0.0074& 0.0682&0.0085& 0.0721&0.0068& 0.0644&0.0021& 0.0362 \\
100&$\hat{\gamma}_Z^{(comp)}(x)$&0.0145& 0.1043&0.0437& 0.1777&0.0658& 0.2188&0.0640& 0.2151&0.0315& 0.1529&0.0600& 0.2108& 0.0684& 0.2209&0.0400&0.1733&0.0138& 0.0982\\
&$\hat{\gamma}_Y(x)$&0.0019& 0.0339&0.0047& 0.0525&0.0084& 0.0719&0.0081& 0.0685&0.0038& 0.0476&0.0067& 0.0640&0.0083& 0.0721& 0.0058& 0.0608&0.0017& 0.0329\\[2ex]

\hline
\multicolumn{20}{c}{ For $\gamma_Y=\gamma_C$ }\\
\hline
&$\hat{\gamma}^{(comp)}_Y(x)$&0.0023& 0.0376&0.0066& 0.0651&0.0116& 0.0847& 0.0093& 0.0744&0.0056& 0.0572&0.0105& 0.0788& 0.01508& 0.0952&0.0094& 0.0733&0.0032& 0.0445\\
100&$\hat{\gamma}_Z^{(comp)}(x)$&0.0024& 0.0378&0.0069& 0.0644&0.0111& 0.0819&0.0101& 0.0781& 0.0057& 0.0580& 0.0121& 0.0850 &  0.0155& 0.1002&  0.0096&0.0764&0.0044& 0.0506\\
&$\hat{\gamma}_Y(x)$&0.0020& 0.0353& 0.0053&0.0560& 0.0088& 0.0734& 0.0076& 0.0670&0.0043& 0.0506&  0.0073& 0.0677&0.0099&0.0750&0.0056& 0.0576&0.0024& 0.0381\\[2ex]
\hline
\multicolumn{20}{c}{\textbf{ For $\gamma_Y>\gamma_C$ }}\\
\hline
&$\hat{\gamma}^{(comp)}_Y(x)$&0.0017& 0.0326&0.0048& 0.0551&0.0092& 0.0748&0.0064& 0.0628&0.0045& 0.0528&0.0078& 0.0697&0.0085&0.0742&0.0058& 0.0599&0.0020& 0.0362 \\
100&$\hat{\gamma}_Z^{(comp)}(x)$&0.0048& 0.0632&0.0132& 0.1024&0.0211& 0.1311&0.0184& 0.1216& 0.0099& 0.0893& 0.0197& 0.1279& 0.0229& 0.1329&0.0147& 0.1074&0.0051& 0.0636\\
&$\hat{\gamma}_Y(x)$&0.0022& 0.0372&0.0079& 0.0672&0.0119& 0.0834&0.0099& 0.0763&0.0053& 0.0564&0.01028& 0.0755& 0.0121& 0.0869& 0.0070&0.0636&0.0029& 0.04036\\[2ex]
\hline
\multicolumn{20}{c}{ For $\gamma_Y<\gamma_C$ }\\
\hline
&$\hat{\gamma}^{(comp)}_Y(x)$&0.0006& 0.0206&0.0018& 0.0345&0.0026& 0.0410& 0.00229& 0.0377& 0.00119& 0.0274&0.0025& 0.0404& 0.0028& 0.0427& 0.0017& 0.03377&  0.0006& 0.0211 \\
300&$\hat{\gamma}_Z^{(comp)}(x)$&0.0101& 0.0873& 0.0277& 0.1452& 0.0468& 0.1923&0.0369& 0.1698& 0.0213& 0.1293& 0.0365& 0.1643&  0.0446& 0.1884&0.0268& 0.1417& 0.0095& 0.0860\\
&$\hat{\gamma}_Y(x)$&0.0005& 0.0181& 0.0015& 0.0303& 0.0025& 0.0402&0.00222& 0.0373&0.00112& 0.0260& 0.0020& 0.0341&0.0024& 0.0394&0.0016&0.0318&0.0005& 0.0182\\[2ex]

\hline
\multicolumn{20}{c}{ For $\gamma_Y=\gamma_C$ }\\
\hline
&$\hat{\gamma}^{(comp)}_Y(x)$&0.00076& 0.0218&0.00222& 0.0363& 0.0032& 0.0452&0.0027& 0.0417& 0.0016& 0.0324&  0.0031& 0.0444& 0.0037& 0.0478& 0.0026& 0.0399& 0.0011& 0.0265\\
300&$\hat{\gamma}_Z^{(comp)}(x)$& 0.00075& 0.0213& 0.00225& 0.0377& 0.0033& 0.0460& 0.0029& 0.0419& 0.0017& 0.033& 0.0034&0.0457&  0.0040& 0.0502& 
0.0027& 0.0410& 0.0009& 0.0239\\
&$\hat{\gamma}_Y(x)$& 0.00059& 0.0193& 0.0017& 0.0334&0.0026& 0.0412& 0.0019& 0.0357&  0.0013& 0.0287& 0.0025&0.0391& 0.0026& 0.0405& 0.0017& 0.0327& 0.0005& 0.0185 \\[2ex]

\hline
\multicolumn{20}{c}{ For $\gamma_Y>\gamma_C$ }\\
\hline
&$\hat{\gamma}^{(comp)}_Y(x)$&0.0006& 0.0195& 0.0018& 0.0348& 0.0034& 0.0453& 0.0025& 0.0410& 0.0015& 0.0305& 0.0032& 0.0452& 0.0036& 0.0470&  0.0025& 0.0400& 0.0009&0.0235 \\
300&$\hat{\gamma}_Z^{(comp)}(x)$&0.0038& 0.0593&  0.0115& 0.1029& 0.0174& 0.1255&  0.0147& 0.1162& 0.0083& 0.0874& 0.0140& 0.1122&  0.0165& 0.1224& 0.0094& 0.0914& 0.0035& 0.0560\\
&$\hat{\gamma}_Y(x)$& 0.0007& 0.0207& 0.0019& 0.0343& 0.0038& 0.0499& 0.0024& 0.0385&  0.0013& 0.0295& 0.0031& 0.0437& 0.0033& 0.0461& 0.0022& 0.0361& 0.00071& 0.0213\\[2ex]
\hline
\multicolumn{20}{c}{ For $\gamma_Y<\gamma_C$ }\\
\hline
&$\hat{\gamma}^{(comp)}_Y(x)$&0.00038& 0.0155& 0.0012& 0.0278&0.0019& 0.0355&0.0017& 0.0338&0.0008& 0.0233&0.0020& 0.0360&0.0019& 0.0349&0.0014& 0.0297& 0.0006& 0.0199\\
500&$\hat{\gamma}_Z^{(comp)}(x)$&0.0091& 0.0831&0.0257& 0.1406&0.0405& 0.1749&0.0359& 0.1657&0.0189& 0.1222&0.0354&0.1644&0.0410& 0.1721&0.0258&0.1369&0.0096& 0.0818\\
&$\hat{\gamma}_Y(x)$&0.00030& 0.0138&0.0010& 0.0251&0.0015& 0.0315&0.0012& 0.0271&0.0006& 0.0205& 0.0012& 0.0282&0.0015& 0.0305&0.0009& 0.0237&0.0003& 0.0145 \\[2ex]

\hline
\multicolumn{20}{c}{ For $\gamma_Y=\gamma_C$ }\\
\hline
&$\hat{\gamma}^{(comp)}_Y(x)$&0.00040& 0.0161&0.00109& 0.0261&0.0023& 0.0389&0.0016& 0.0327&0.0008& 0.0234&  0.0016& 0.0314& 0.0022& 0.0373&0.0013& 0.0284&0.0006& 0.0200 \\
500&$\hat{\gamma}_Z^{(comp)}(x)$&0.00041& 0.0158&0.00109&0.0260&0.0019& 0.0351&0.0015& 0.0315&0.0009& 0.0244&0.0015& 0.0307&0.0023& 0.0380&0.0013& 0.0297&0.0005& 0.0188\\
&$\hat{\gamma}_Y(x)$&0.00033& 0.0145&0.00102& 0.02549&0.0014& 0.0308&0.0012& 0.0276&0.0007& 0.0220&0.0012& 0.0274&0.0016&0.0315& 0.0009& 0.0248&0.0003& 0.0151\\[2ex]
\hline
\multicolumn{20}{c}{ For $\gamma_Y>\gamma_C$ }\\
\hline
&$\hat{\gamma}^{(comp)}_Y(x)$&0.0004& 0.0162& 0.0010& 0.0247& 0.0019& 0.0346&0.0013& 0.0295& 0.0009& 0.0241&0.0015& 0.0315&0.0020& 0.0365& 0.0012& 0.0276& 0.0004& 0.0167 \\
500&$\hat{\gamma}_Z^{(comp)}(x)$&0.0040& 0.0624&0.011& 0.1057&0.0183& 0.1318& 0.0151& 0.1199& 0.0085& 0.0900&0.0148& 0.1187&0.0181& 0.1309&0.0105& 0.0998&0.0038& 0.0600\\
&$\hat{\gamma}_Y(x)$&0.0004& 0.0162&0.0012& 0.0277& 0.0021& 0.0359&0.0018& 0.0345&0.0009& 0.0240& 0.0018& 0.0342&0.0017& 0.0339&0.0014& 0.0296&0.0004& 0.0166\\[2ex]

\hline
\end{tabular}}
\end{sidewaystable}
\FloatBarrier

\FloatBarrier
 
\FloatBarrier

\FloatBarrier
 
\begin{sidewaystable}[ht!]
\caption{Simulation results for the estimator of $q(\alpha_n|\cdot)$ where $\alpha_n=1/1000$: empirical MSE and empirical MAE for $N = 500$ replications.}
  \label{tab:2}
\centering
\scalebox{.63}{\begin{tabular}{c|c|cc|cc|cc|cc|cc|cc|cc|cc|cc}
\hline
&estimator&\multicolumn{2}{|c|}{ $q(\alpha_n|0.1)=0.2114$ }&\multicolumn{2}{|c|}{  $q(\alpha_n|0.2)=0.0735$ }&\multicolumn{2}{|c}{ $q(\alpha_n|0.3)=0.0357$}&\multicolumn{2}{|c|}{ $q(\alpha_n|0.4)=0.0480$ }&\multicolumn{2}{|c|}{  $q(\alpha_n|0.5)=0.1023$ }&\multicolumn{2}{|c}{ $q(\alpha_n|0.6)=0.0480$}&\multicolumn{2}{|c|}{$q(\alpha_n|0.7)=0.0357$ }&\multicolumn{2}{|c|}{  $q(\alpha_n|0.8)=0.0735$  }&\multicolumn{2}{|c}{ $q(\alpha_n|0.9)=0.2114$}\\
\hline
$n$&&MSE&MAE&MSE&MAE&MSE&MAE&MSE&MAE&MSE&MAE&MSE&MAE&MSE&MAE&MSE&MAE&MSE&MAE\\
\hline
\multicolumn{20}{c}{ For $\gamma_Y<\gamma_C$ }\\
\hline
&$\hat{\gamma}^{(comp)}_Y(x)$&0.0037& 0.0453& 0.0012& 0.0256&0.0004& 0.0152&0.0007& 0.0197&0.0016& 0.0313&0.0006& 0.0193&0.00039& 0.0145&0.0011& 0.0259&0.0037& 0.0449\\
100&$\hat{\gamma}_Z^{(comp)}(x)$&0.0093& 0.0813&0.0018& 0.0384&0.0005& 0.0206&0.0009& 0.0262&0.0032& 0.0499&0.0009& 0.0273&0.00054& 0.0206&0.0019& 0.0381&0.0090& 0.0796\\
&$\hat{\gamma}_Y(x)$&0.0033& 0.0425&0.0008& 0.0220&0.0003& 0.0144&0.0006& 0.0182&0.0013& 0.0286&0.0006& 0.0181&0.00035&0.0139&0.0010& 0.0243&0.0031& 0.0421\\[2ex]

\hline
\multicolumn{20}{c}{ For $\gamma_Y=\gamma_C$ }\\
\hline
&$\hat{\gamma}^{(comp)}_Y(x)$&0.0040& 0.0496&0.00108& 0.0263&0.00048& 0.0158&0.00059& 0.0193&0.0019& 0.0346&0.0007& 0.0212&0.00048& 0.0169&0.0012& 0.0279&0.0044& 0.0545  \\
100&$\hat{\gamma}_Z^{(comp)}(x)$&0.0037& 0.0484&0.00105& 0.0258&0.00046& 0.0164&0.00059& 0.0198&0.0016& 0.0337&0.0006& 0.0201& 0.00048&0.0170&0.0011& 0.0278&0.0047& 0.0560\\
&$\hat{\gamma}_Y(x)$&0.0034& 0.0456& 0.00103& 0.0265& 0.00039& 0.0157&0.00070& 0.0199&0.0017& 0.0324&0.0007& 0.0201&0.00044&0.0160&0.0009& 0.0252&0.0034& 0.0455\\[2ex]
\hline
\multicolumn{20}{c}{ For $\gamma_Y>\gamma_C$ }\\
\hline
&$\hat{\gamma}^{(comp)}_Y(x)$& 0.0038& 0.0485&0.0012&9 0.0275&0.0006& 0.0182&0.0008& 0.0214& 0.0020& 0.0357&0.0009& 0.0231&0.0006& 0.0189&0.0016& 0.0308&0.0044& 0.0519\\
100&$\hat{\gamma}_Z^{(comp)}(x)$&0.0151& 0.1025&0.0077& 0.0697&0.0040& 0.0497&0.0056& 0.0595&.0096& 0.0806&0.0066& 0.0663&0.0047& 0.0531&0.0096& 0.0776&0.0165& 0.1071\\
&$\hat{\gamma}_Y(x)$&0.0039& 0.0505&0.0012&0.0283&0.0006& 0.0186& 0.0007& 0.0215&0.0018& 0.0337&0.0008& 0.0220&0.0005& 0.0183&0.0015& 0.0306&0.0046& 0.0531\\[2ex]
\hline
\multicolumn{20}{c}{ For $\gamma_Y<\gamma_C$ }\\
\hline
&$\hat{\gamma}^{(comp)}_Y(x)$&0.0015& 0.0275&0.0004& 0.0155& 0.0001& 0.0091&0.0002& 0.0107&0.0005& 0.0169&0.0002& 0.0115&0.00015& 0.0094&0.0004& 0.0150&0.0013& 0.0278\\
300&$\hat{\gamma}_Z^{(comp)}(x)$& 0.0067& 0.0723& 0.0015& 0.0353&0.0004&0.0199&0.0007& 0.0251&0.0027& 0.0471&0.0007&0.0246&0.0004& 0.0197&0.0014& 0.0338&0.0066& 0.0714\\
&$\hat{\gamma}_Y(x)$& 0.0010& 0.0236&0.0003& 0.0136&0.0001& 0.0088&0.0002& 0.0110&0.0005& 0.0172&0.0001& 0.0101&0.0001& 0.0084&0.0003& 0.0145&0.0009& 0.0242\\[2ex]

\hline
\multicolumn{20}{c}{ For $\gamma_Y=\gamma_C$ }\\
\hline
&$\hat{\gamma}^{(comp)}_Y(x)$&0.0012& 0.0272& 0.00041& 0.0160&0.0001& 0.0096& 0.000198& 0.0115&0.0006& 0.0194&0.00023& 0.0122&0.00013& 0.0093&0.00040& 0.0162&0.0015& 0.0326\\
300&$\hat{\gamma}_Z^{(comp)}(x)$& 0.0012& 0.0272& 0.00044& 0.0165& 0.0001& 0.0093&0.000194& 0.0113&0.00056& 0.0192&0.00023& 0.0126& 0.00015& 0.0102&0.00044& 0.0173&0.0014&0.0303\\
&$\hat{\gamma}_Y(x)$& 0.0011& 0.0258&0.00040& 0.0154&0.0001& 0.0093&0.0002& 0.0110& 0.00051& 0.0180&0.00021& 0.0114&0.00013& 0.0089& 0.00041& 0.0157&0.0010& 0.0253\\[2ex]

\hline
\multicolumn{20}{c}{ For $\gamma_Y>\gamma_C$ }\\
\hline
&$\hat{\gamma}^{(comp)}_Y(x)$&0.0011&0.0264&0.00040& 0.0160& 0.00016& 0.0098&0.00023& 0.0121&0.00061& 0.0194& 0.00023& 0.0126&0.00013& 0.0092&0.00044& 0.0170& 0.0014& 0.0303
 \\
300&$\hat{\gamma}_Z^{(comp)}(x)$&0.0104& 0.0936&0.0054& 0.0671&0.00236& 0.0429&0.0033& 0.0519& 0.0067& 0.0748& 0.0031& 0.0494&0.0022& 0.0415&0.0041& 0.0571& 0.0094& 0.0866\\
&$\hat{\gamma}_Y(x)$&0.0012& 0.0286&  0.00043& 0.0163&0.00019& 0.011& 0.00026& 0.0123&0.00064& 0.0198& 0.00030& 0.0132&0.00017& 0.0103& 0.00045& 0.0169&0.0013& 0.0293\\[2ex]
\hline
\multicolumn{20}{c}{ For $\gamma_Y<\gamma_C$ }\\
\hline
&$\hat{\gamma}^{(comp)}_Y(x)$&0.0007& 0.0196& 0.0002& 0.0122& 9.6e-05& 0.0074& 0.00013& 0.0092&0.0003&9 0.0138&0.00014& 0.0093&8.6e-05& 0.0072&0.00029& 0.0132&0.0009& 0.0239\\
500&$\hat{\gamma}_Z^{(comp)}(x)$& 0.0066& 0.0731&0.0015& 0.0361&4.7e-04& 0.0200&0.00079& 0.0260&0.0025& 0.0451&0.00077& 0.0255&4.2e-04& 0.0183&0.00136& 0.0322&0.0063& 0.0706\\
&$\hat{\gamma}_Y(x)$&0.0006& 0.0196& 0.0002& 0.0119&8.7e-05& 0.0072&0.00011& 0.0083&0.0003& 0.0140& 0.00011& 0.0082&7.4e-05& 0.0066&0.00021& 0.0113&0.0007& 0.0195\\[2ex]

\hline
\multicolumn{20}{c}{ For $\gamma_Y=\gamma_C$ }\\
\hline
&$\hat{\gamma}^{(comp)}_Y(x)$&0.00069& 0.0210&0.0002& 0.0118&1.0e-04& 0.0084&0.00013& 0.0092& 0.0003& 0.0146& 0.00012& 0.0088&1.0e-04& 0.0081&0.00025& 0.0125&0.0009&0.0252\\
500&$\hat{\gamma}_Z^{(comp)}(x)$&0.00072& 0.0208&0.0002& 0.0118&9.0e-05& 0.0076&0.00012& 0.0090& 0.0003& 0.0152&0.00012& 0.0087&8.9e-05& 0.0077&0.00025& 0.0128&0.0009& 0.0236\\
&$\hat{\gamma}_Y(x)$&0.00062& 0.0198&0.0002& 0.0120&8.1e-05& 0.0071&0.00012& 0.0086&0.0003& 0.0146& 0.00011& 0.0082& 8.2e-05& 0.0070&0.00020& 0.0114&0.0006& 0.0209\\[2ex]
\hline
\multicolumn{20}{c}{ For $\gamma_Y>\gamma_C$ }\\
\hline
&$\hat{\gamma}^{(comp)}_Y(x)$&0.0007& 0.0217&0.0002& 0.0115& 9.71e-05& 0.0077&0.0001& 0.0092&0.0003& 0.0149&0.0001& 0.0095& 1.06e-04& 0.0080&0.0002& 0.0124&0.0007& 0.0226\\

500&$\hat{\gamma}_Z^{(comp)}(x)$&0.0104& 0.0964&0.0051& 0.0666& 2.1e-03& 0.0445&0.0032& 0.0530&0.0065& 0.0753&0.0031& 0.0523&2.4e-03 &0.0449& 0.0045& 0.0633&0.0102& 0.0945\\

&$\hat{\gamma}_Y(x)$&0.0008& 0.0225&0.0002& 0.0132&1.05e-04& 0.0078&0.0001& 0.0104& 0.0003& 0.0153& 0.0001& 0.0104&9.4e-05 &0.0075&0.0003& 0.0138&  0.0007&.0218\\[2ex]

\hline
\end{tabular}}
\end{sidewaystable}
\FloatBarrier

\FloatBarrier
 
\FloatBarrier

\FloatBarrier

\section{Illustration on real data}\label{section5}
The  main goal of this section is  to illustrate our methodology on real dataset of men suffering from a larynx cancer. According to the literature,  this dataset was  previously  analyzed  by \cite{worms2019estimation}, where  the authors considered $Z$ as  the time to death if $\delta =1$ or on-study time if $\delta =0$ in months. The comparison of our results  with the existence results in literature  will be considered.  The dataset contains $n=90$ male patients diagnosed with  a larynx cancer, the dataset is  available and  fully presented in \cite{book:1298616}. \cite{gomes2011estimation} introduced  the aspect of the estimation  of extreme values using  this datasets. The information on each patient includes the time on study in month, the age at diagnosis, the age of diagnosed, stage  of disease as indicator which equals $1$ if the patient died and 0 otherwise. In this dataset $50$ patients died; the other survival times are right-censored. \cite{worms2019estimation}  estimated the extreme value index $\gamma_Y$ and  Weissman extreme quantile (with $\alpha_n=0.05$) of the (unconditional) distribution $\bar{F}_y(\cdot)$ of the survival time $Y$, then they considered  $k_n=37$  which gives $\hat{x}_p(0.05)= 22$ as presented in \cite{worms2019estimation}.   

\begin{figure}[ht!]
\centering
  \includegraphics[width=8cm,height=8cm]{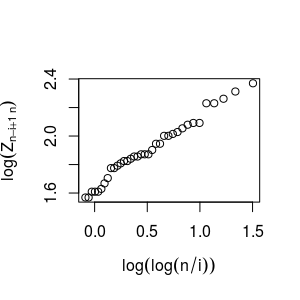}
\caption{Weibull Quantile-Quantile  for $k_n=54$ with dataset of men suffering from a larynx cancer .} 
\label{fig:8}
\end{figure}

\noindent The Weibull quantile-quantile plots for the data is considered, we  plot the  points $(\log \log(n/i)$, $\log(Z_{n-i+1,n}))$,  for $ i=1,\cdots,k_n$ , for well chosen  value of $k_n$ the graphical presentation is illustrated in Figure \ref{fig:8} to support the assumption  that a Weibull-tail distribution is a possible  to the dataset of men suffering from a larynx cancer. The  remaining challenge is the right choice  of the $k_n$ and $h$. 
In this  section, to determine optimum values of $h$ and   threshold  excess $k$,  we use  the same methods  as mentioned  in Section \ref{design}.\\
\noindent By  taking into consideration  of the presence of covariate  and using the aforementioned  strategy  to determine the  appropriate value of $k$, we get $k=54$. 
\FloatBarrier
We rerun these data, taking account of the age at diagnosis denoted by x in what follows.
 For  illustration of our proposed  estimator, We estimate the quantile $q(0.05|x)$ of order     $1-0.05$ of the conditional distribution of $Z$ given $x$ for the case $ x = 65$ (median of the $(x,\cdots,x_n)$, $x = 54.20 \; ( median-sd(x_1,\cdots,x_n))$ and $x = 75.80\; ( median + sd (x_1,\cdots,x_n ))$  where $sd ( x_1,\cdots,x_n$ is the empirical standard deviation of $x_1,\cdots,x_n$.
The results are presented in  Table \ref{tab:3}.

 \begin{table}[ht!]
\caption{Estimation of the Weibull-tail coefficient and conditional extreme quantile, first column: $median(x)-sd(x)$, second  column: $median(x)$ and third column: $median(x)+sd(x)$.}
\centering
\begin{tabular}{c|cccc|}
\hline
x& 54.20&   65& 75.80\\
\hline
\multicolumn{4}{c}{ For $k=54$ }\\
\hline
$\hat{\gamma}_Y(x)$&0.8219978 & 0.8225810 & 0.8210728\\
$\hat{q}_w(0.05|x)$&17.6013366& 17.6222551& 17.5682097\\[2ex]
\hline
\multicolumn{4}{c}{ For $k=37$ }\\ \hline
$\hat{\gamma}_Y(x)$&1.0164362  &1.017575 & 1.0146309\\
$\hat{q}_w(0.05|x)$&23.1668062& 23.209382& 23.0994800\\[2ex]
\hline
\end{tabular}
  \label{tab:3}
\end{table}
\FloatBarrier
Finally, from Table \ref{tab:3}, we observe that for every $x$ the estimates \eqref{EqWquantile} are smaller than the unconditional estimate obtained by \cite{worms2019estimation}, thus providing less optimistic perspectives for being infected. But we expect our results to be more representative of the real chance of  being survive of these larynx cancer patients, since our analysis takes account of the influent variable "age at diagnosis".

\section{Conclusion and Perspectives }\label{section6}

In this paper, the estimation of the Weibull tail coefficient and extreme quantiles of a Weibull tail type  distribution when some covariate information is available and the data are randomly right-censored are considered. The proposed  estimators for conditional Weibull tail coefficient and  conditional Weissman quantile are derived. The parameter estimation method is proposed to prove its
performance. We assessed their finite-sample performance via simulations. A comparison with two  estimation strategies has been provided. Our intensive simulation study shows that the
proposed estimators are competitive in all scenario as  the sample size  becomes large enough. 

A future possible work would be to exploit our  proposed  methodology  in presence  of functional random covariate. 
Some additional research is needed to  establish asymptotic normality of  our proposed estimators.
\section*{References}
\bibliography{Rutikanga_Diop}

\begin{thebibliography}{24}
\expandafter\ifx\csname natexlab\endcsname\relax\def\natexlab#1{#1}\fi
\providecommand{\url}[1]{\texttt{#1}}
\providecommand{\href}[2]{#2}
\providecommand{\path}[1]{#1}
\providecommand{\DOIprefix}{doi:}
\providecommand{\ArXivprefix}{arXiv:}
\providecommand{\URLprefix}{URL: }
\providecommand{\Pubmedprefix}{pmid:}
\providecommand{\doi}[1]{\href{http://dx.doi.org/#1}{\path{#1}}}
\providecommand{\Pubmed}[1]{\href{pmid:#1}{\path{#1}}}
\providecommand{\bibinfo}[2]{#2}
\ifx\xfnm\relax \def\xfnm[#1]{\unskip,\space#1}\fi
\bibitem[{Beirlant et~al.(1996)Beirlant, Teugels \&
  Vynckier}]{beirlant1996practical}
\bibinfo{author}{Beirlant, J.}, \bibinfo{author}{Teugels, J.~L.}, \&
  \bibinfo{author}{Vynckier, P.} (\bibinfo{year}{1996}).
\newblock {\it \bibinfo{title}{Practical analysis of extreme values}\/}
  volume~\bibinfo{volume}{50}.
\newblock \bibinfo{publisher}{Leuven University Press Leuven}.
\bibitem[{Beirlant et~al.(2004)Beirlant, Yuri, Jozef, Johan, Daniel \&
  Chris}]{beirlant2004estimation}
\bibinfo{author}{Beirlant, J.}, \bibinfo{author}{Yuri, G.},
  \bibinfo{author}{Jozef, T.}, \bibinfo{author}{Johan, S.},
  \bibinfo{author}{Daniel, D.~W.}, \& \bibinfo{author}{Chris, F.}
  (\bibinfo{year}{2004}).
\newblock \bibinfo{title}{Statistics of extremes}.
\newblock {\it \bibinfo{journal}{John Wiley \& Sons Ltd}\/},  {\it
  \bibinfo{volume}{10}\/}, \bibinfo{pages}{151--174}.
\bibitem[{Berred(1991)}]{berred1991record}
\bibinfo{author}{Berred, M.} (\bibinfo{year}{1991}).
\newblock \bibinfo{title}{Record values and the estimation of the weibull
  tail-coefficient}.
\newblock {\it \bibinfo{journal}{Comptes rendus de l'Acad{\'e}mie des sciences.
  S{\'e}rie 1, Math{\'e}matique}\/},  {\it \bibinfo{volume}{312}\/},
  \bibinfo{pages}{943--946}.
\bibitem[{Cabras \& Castellanos(2011)}]{cabras_castellanos_2011}
\bibinfo{author}{Cabras, S.}, \& \bibinfo{author}{Castellanos, M.~E.}
  (\bibinfo{year}{2011}).
\newblock \bibinfo{title}{A bayesian approach for estimating extreme quantiles
  under a semiparametric mixture model}.
\newblock {\it \bibinfo{journal}{ASTIN Bulletin}\/},  {\it
  \bibinfo{volume}{41}\/}, \bibinfo{pages}{87–106}.
  \DOIprefix\doi{10.2143/AST.41.1.2084387}.
\bibitem[{Coles \& Powell(1996)}]{coles1996bayesian}
\bibinfo{author}{Coles, S.~G.}, \& \bibinfo{author}{Powell, E.~A.}
  (\bibinfo{year}{1996}).
\newblock \bibinfo{title}{Bayesian methods in extreme value modelling: a review
  and new developments}.
\newblock {\it \bibinfo{journal}{International Statistical Review/Revue
  Internationale de Statistique}\/},  (pp. \bibinfo{pages}{119--136}).
\bibitem[{Diebolt et~al.(2008)Diebolt, Gardes, Girard \&
  Guillou}]{diebolt2008bias}
\bibinfo{author}{Diebolt, J.}, \bibinfo{author}{Gardes, L.},
  \bibinfo{author}{Girard, S.}, \& \bibinfo{author}{Guillou, A.}
  (\bibinfo{year}{2008}).
\newblock \bibinfo{title}{Bias-reduced extreme quantile estimators of weibull
  tail-distributions}.
\newblock {\it \bibinfo{journal}{Journal of Statistical Planning and
  Inference}\/},  {\it \bibinfo{volume}{138}\/}, \bibinfo{pages}{1389--1401}.
\bibitem[{Fisher \& Tippett(1928)}]{fisher_tippett_1928}
\bibinfo{author}{Fisher, R.~A.}, \& \bibinfo{author}{Tippett, L. H.~C.}
  (\bibinfo{year}{1928}).
\newblock \bibinfo{title}{Limiting forms of the frequency distribution of the
  largest or smallest member of a sample}.
\newblock {\it \bibinfo{journal}{Mathematical Proceedings of the Cambridge
  Philosophical Society}\/},  {\it \bibinfo{volume}{24}\/},
  \bibinfo{pages}{180–190}. \DOIprefix\doi{10.1017/S0305004100015681}.
\bibitem[{Gardes \& Girard(2012)}]{gardes2012functional}
\bibinfo{author}{Gardes, L.}, \& \bibinfo{author}{Girard, S.}
  (\bibinfo{year}{2012}).
\newblock \bibinfo{title}{Functional kernel estimators of large conditional
  quantiles}.
\newblock {\it \bibinfo{journal}{Electronic Journal of Statistics}\/},  {\it
  \bibinfo{volume}{6}\/}, \bibinfo{pages}{1715--1744}.
\bibitem[{Gardes \& Girard(2016)}]{gardes2016estimation}
\bibinfo{author}{Gardes, L.}, \& \bibinfo{author}{Girard, S.}
  (\bibinfo{year}{2016}).
\newblock \bibinfo{title}{On the estimation of the functional weibull
  tail-coefficient}.
\newblock {\it \bibinfo{journal}{Journal of Multivariate Analysis}\/},  {\it
  \bibinfo{volume}{146}\/}, \bibinfo{pages}{29--45}.
\bibitem[{Goegebeur et~al.(2014{\natexlab{a}})Goegebeur, Guillou \&
  Osmann}]{goegebeur2014local}
\bibinfo{author}{Goegebeur, Y.}, \bibinfo{author}{Guillou, A.}, \&
  \bibinfo{author}{Osmann, M.} (\bibinfo{year}{2014}{\natexlab{a}}).
\newblock \bibinfo{title}{A local moment type estimator for the extreme value
  index in regression with random covariates}.
\newblock {\it \bibinfo{journal}{Canadian Journal of Statistics}\/},  {\it
  \bibinfo{volume}{42}\/}, \bibinfo{pages}{487--507}.
\bibitem[{Goegebeur et~al.(2014{\natexlab{b}})Goegebeur, Guillou \&
  Schorgen}]{goegebeur2014nonparametric}
\bibinfo{author}{Goegebeur, Y.}, \bibinfo{author}{Guillou, A.}, \&
  \bibinfo{author}{Schorgen, A.} (\bibinfo{year}{2014}{\natexlab{b}}).
\newblock \bibinfo{title}{Nonparametric regression estimation of conditional
  tails: the random covariate case}.
\newblock {\it \bibinfo{journal}{Statistics}\/},  {\it \bibinfo{volume}{48}\/},
  \bibinfo{pages}{732--755}.
\bibitem[{Gomes \& Neves(2011)}]{gomes2011estimation}
\bibinfo{author}{Gomes, M.~I.}, \& \bibinfo{author}{Neves, M.~M.}
  (\bibinfo{year}{2011}).
\newblock \bibinfo{title}{Estimation of the extreme value index for randomly
  censored data}.
\newblock {\it \bibinfo{journal}{Biometrical Letters}\/},  {\it
  \bibinfo{volume}{48}\/}, \bibinfo{pages}{1--22}.
\bibitem[{John P.~Klein(2005)}]{book:1298616}
\bibinfo{author}{John P.~Klein, M. L.~M.} (\bibinfo{year}{2005}).
\newblock {\it \bibinfo{title}{Survival Analysis: Techniques for Censored and
  Truncated Data}\/}.
\newblock Statistics for Biology and Health (\bibinfo{edition}{2nd} ed.).
\newblock \bibinfo{publisher}{Springer}.
\newblock \URLprefix
  \url{http://gen.lib.rus.ec/book/index.php?md5=8a70c414ede17984621958bdc45688e0}.
\bibitem[{Matthys et~al.(2004)Matthys, Delafosse, Guillou \&
  Beirlant}]{matthys2004estimating}
\bibinfo{author}{Matthys, G.}, \bibinfo{author}{Delafosse, E.},
  \bibinfo{author}{Guillou, A.}, \& \bibinfo{author}{Beirlant, J.}
  (\bibinfo{year}{2004}).
\newblock \bibinfo{title}{Estimating catastrophic quantile levels for
  heavy-tailed distributions}.
\newblock {\it \bibinfo{journal}{Insurance: Mathematics and Economics}\/},
  {\it \bibinfo{volume}{34}\/}, \bibinfo{pages}{517--537}.
\bibitem[{Nadaraya(1964)}]{nadaraya1964estimating}
\bibinfo{author}{Nadaraya, E.~A.} (\bibinfo{year}{1964}).
\newblock \bibinfo{title}{On estimating regression}.
\newblock {\it \bibinfo{journal}{Theory of Probability \& Its Applications}\/},
   {\it \bibinfo{volume}{9}\/}, \bibinfo{pages}{141--142}.
\bibitem[{Ndao(2015)}]{ndao2015modelisation}
\bibinfo{author}{Ndao, P.} (\bibinfo{year}{2015}).
\newblock {\it \bibinfo{title}{Mod{\'e}lisation de valeurs extr{\^e}mes
  conditionnelles en pr{\'e}sence de censure}\/}.
\newblock \bibinfo{publisher}{PhD thesis, Universit{\'e} Gaston Berger de
  Saint-Louis}.
\bibitem[{Ndao et~al.(2014)Ndao, Diop \& Dupuy}]{ndao2014nonparametric}
\bibinfo{author}{Ndao, P.}, \bibinfo{author}{Diop, A.}, \&
  \bibinfo{author}{Dupuy, J.-F.} (\bibinfo{year}{2014}).
\newblock \bibinfo{title}{Nonparametric estimation of the conditional tail
  index and extreme quantiles under random censoring}.
\newblock {\it \bibinfo{journal}{Computational Statistics \& Data Analysis}\/},
   {\it \bibinfo{volume}{79}\/}, \bibinfo{pages}{63--79}.
\bibitem[{Ndao et~al.(2016)Ndao, Diop \& Dupuy}]{ndao2016nonparametric}
\bibinfo{author}{Ndao, P.}, \bibinfo{author}{Diop, A.}, \&
  \bibinfo{author}{Dupuy, J.-F.} (\bibinfo{year}{2016}).
\newblock \bibinfo{title}{Nonparametric estimation of the conditional
  extreme-value index with random covariates and censoring}.
\newblock {\it \bibinfo{journal}{Journal of Statistical Planning and
  Inference}\/},  {\it \bibinfo{volume}{168}\/}, \bibinfo{pages}{20--37}.
\bibitem[{Stephenson \& Tawn(2004)}]{stephenson2004bayesian}
\bibinfo{author}{Stephenson, A.}, \& \bibinfo{author}{Tawn, J.}
  (\bibinfo{year}{2004}).
\newblock \bibinfo{title}{Bayesian inference for extremes: accounting for the
  three extremal types}.
\newblock {\it \bibinfo{journal}{Extremes}\/},  {\it \bibinfo{volume}{7}\/},
  \bibinfo{pages}{291--307}.
\bibitem[{Stupfler(2016)}]{stupfler2016estimating}
\bibinfo{author}{Stupfler, G.} (\bibinfo{year}{2016}).
\newblock \bibinfo{title}{Estimating the conditional extreme-value index under
  random right-censoring}.
\newblock {\it \bibinfo{journal}{Journal of Multivariate Analysis}\/},  {\it
  \bibinfo{volume}{144}\/}, \bibinfo{pages}{1--24}.
\bibitem[{Watson(1964)}]{watson1964smooth}
\bibinfo{author}{Watson, G.~S.} (\bibinfo{year}{1964}).
\newblock \bibinfo{title}{Smooth regression analysis}.
\newblock {\it \bibinfo{journal}{Sankhy{\=a}: The Indian Journal of Statistics,
  Series A}\/},  (pp. \bibinfo{pages}{359--372}).
\bibitem[{de~Wet et~al.(2016)de~Wet, Goegebeur, Guillou \&
  Osmann}]{de2016kernel}
\bibinfo{author}{de~Wet, T.}, \bibinfo{author}{Goegebeur, Y.},
  \bibinfo{author}{Guillou, A.}, \& \bibinfo{author}{Osmann, M.}
  (\bibinfo{year}{2016}).
\newblock \bibinfo{title}{Kernel regression with weibull-type tails}.
\newblock {\it \bibinfo{journal}{Annals of the Institute of Statistical
  Mathematics}\/},  {\it \bibinfo{volume}{68}\/}, \bibinfo{pages}{1135--1162}.
\bibitem[{Worms \& Worms(2014)}]{worms2014new}
\bibinfo{author}{Worms, J.}, \& \bibinfo{author}{Worms, R.}
  (\bibinfo{year}{2014}).
\newblock \bibinfo{title}{New estimators of the extreme value index under
  random right censoring, for heavy-tailed distributions}.
\newblock {\it \bibinfo{journal}{Extremes}\/},  {\it \bibinfo{volume}{17}\/},
  \bibinfo{pages}{337--358}.
\bibitem[{Worms \& Worms(2019)}]{worms2019estimation}
\bibinfo{author}{Worms, J.}, \& \bibinfo{author}{Worms, R.}
  (\bibinfo{year}{2019}).
\newblock \bibinfo{title}{Estimation of extremes for weibull-tail distributions
  in the presence of random censoring}.
\newblock {\it \bibinfo{journal}{Extremes}\/},  {\it \bibinfo{volume}{22}\/},
  \bibinfo{pages}{667--704}.

\end{thebibliography}

\end{document}